\documentclass[conference]{IEEEtran}
\ifCLASSINFOpdf
\else
\fi
\usepackage{array}
\usepackage{outlines}
\usepackage{cite}
\usepackage{url}
\usepackage{amsmath,amssymb,amsfonts}
\usepackage{graphicx}
\usepackage{textcomp}
\usepackage{listings}
\usepackage{xcolor}
\usepackage{wrapfig}
\usepackage{multirow}
\usepackage{booktabs}
\usepackage{multirow}
\usepackage{float}
\usepackage{placeins}
\usepackage{eufrak}
 \usepackage{tikz} 
 
\usepackage{annotate-equations}
 
\usepackage{framed}
\usepackage[font=small,skip=1pt]{caption}
\usepackage{subcaption}
\usepackage{graphicx,xcolor} 
\usepackage[framemethod=tikz]{mdframed}
\usepackage{algorithm}
\usepackage{algorithmic}
\usepackage{lipsum}
\usepackage{tikz}
\usepackage{nicefrac}
\usepackage[breaklinks]{hyperref}
\usepackage{breakurl}
\usepackage{tikz}
\usepackage{xcolor}
\usepackage{threeparttable}
\newcommand*\circled[1]{\tikz[baseline=(char.base)]{
            \node[shape=circle,fill,inner sep=2pt] (char) {\textcolor{white}{#1}};}}

\definecolor{codegreen}{rgb}{0,0.6,0}
\definecolor{codegray}{rgb}{0.5,0.5,0.5}
\definecolor{codepurple}{rgb}{0.58,0,0.82}
\definecolor{backcolour}{RGB}{242, 242, 251}

\lstdefinestyle{mystyle}{
    backgroundcolor=\color{backcolour},   
    commentstyle=\color{codegreen},
    keywordstyle=\color{magenta},
    numberstyle=\tiny\color{codegray},
    stringstyle=\color{codepurple},
    basicstyle=\ttfamily\footnotesize,
    breakatwhitespace=false,         
    breaklines=true,                 
    captionpos=b,                    
    keepspaces=true,                 
    numbers=left,                    
    numbersep=5pt,                  
    showspaces=false,                
    showstringspaces=false,
    showtabs=false,                  
    tabsize=2
}
\begin{document}
\bstctlcite{IEEEexample:BSTcontrol}
%
\title{A Software-Hardware Co-Optimized Toolkit for Deep Reinforcement Learning on\\ Heterogeneous Platforms}

\author{
    \IEEEauthorblockN{
        Yuan Meng$^1$, Michael Kinsner$^2$, Deshanand Singh$^2$, Mahesh A Iyer$^2$, Viktor Prasanna$^1$
    }
    \IEEEauthorblockA{
        \textit{$^1$Ming Hsieh Department of Electrical and Computer Engineering, University of Southern California} \\
        \textit{$^2$Intel Corporation}\\
        Contact: \{ymeng643,prasanna\}@usc.edu
    }
}

\maketitle


\begin{abstract}
Deep Reinforcement Learning (DRL) is vital in various AI applications.
DRL algorithms comprise diverse compute kernels, which may not be simultaneously optimized using a homogeneous architecture.
However, even with available heterogeneous architectures, optimizing DRL performance remains a challenge due to the complexity of hardware and programming models employed in modern data centers.
To address this, we introduce PEARL, a toolkit for composing parallel DRL systems on heterogeneous platforms consisting of general-purpose processors (CPUs) and accelerators (GPUs, FPGAs).
Our innovations include:
1. A general training protocol agnostic of the underlying hardware, enabling portable implementations across various processors and accelerators. 
2. Incorporation of DRL-specific scheduling optimizations within the protocol, facilitating parallelized training and enhancing the overall system performance.
3. High-level API for productive development using the toolkit.
4. Automatic optimization of DRL task-to-device assignments 
through performance estimation, supporting 
various optimization metrics including throughput and power efficiency.

We showcase our toolkit through experimentation with two widely used DRL algorithms, DQN and DDPG, on two diverse heterogeneous platforms.
The generated implementations outperform state-of-the-art libraries for CPU-GPU platforms by throughput improvements of up to 2.1$\times$ and power efficiency improvements of up to 3.4$\times$.
\end{abstract}


%
\IEEEpeerreviewmaketitle

\section{Introduction}

Deep Reinforcement Learning (DRL) is extensively applied in various domains, including robotics, surveillance, etc. \cite{chatzilygeroudis2017black, vinyals2019alphastar}. 
Most DRL algorithms involve three collaborative compute kernels: policy execution, training, and dataset management. In policy execution, parallel Actors gather data through inference on the policy, interact with the environment, and deposit the data into a Prioritized Replay Buffer for dataset storage. In training, a centralized Learner samples data from the Prioritized Replay Buffer to update the policy model. The dataset management within the Prioritized Replay Buffer is facilitated by a sum tree data structure storing data priorities \cite{zhang2021parallel}.

DRL training is highly time consuming. Due to the distinct compute kernels in DRL that may not be efficiently optimized using a homogeneous architecture, there has been a need and growing trend in using heterogeneous architectures to map and accelerate DRL algorithms \cite{ray_rllib,cho2019fa3c,meng2020accelerating}. However, even with access to heterogeneous resources, DRL application developers still face several challenges:
(a). \textit{Sub-optimal performance}: 
DRL's distinct components require careful placement and scheduling onto heterogeneous devices based on both computational and hardware characteristics. Sub-optimal placement and scheduling can lead to under-utilization of heterogeneous resources, resulting in missed opportunities for performance improvement.
(b). \textit{Lack of portability across different platforms}: 
The optimal DRL primitive-to-hardware assignments can change based on varying algorithms and platforms. Consistently achieving high performance implementations requires portable solutions that can map DRL onto various devices, but existing frameworks lack such flexibility.
(c). \textit{Low development productivity}: The growing diversity of heterogeneous resources in data centers \cite{devcloud,yasar2022pgabb,van2012accelerating} have increased the need for hardware optimizations and bridging between different programming models. This significantly increases the required learning effort and programming time for application developers.

In this work, we address the above challenges by proposing PEARL, a toolkit that enhances the performance, productivity, and portability \cite{pennycook2021navigating} of DRL system development on heterogeneous platforms. 
Our toolkit provides end-to-end support for application developers, allowing them to specify algorithms and benchmarks on the front end, while the toolkit generates the low-level parallel designs on the back end.
oneAPI \cite{oneapi} is a tool flow for creating heterogeneous applications across processors and accelerator architectures using Data Parallel C++ \cite{dpcpp}, which is based on the SYCL open standard \cite{sycl}. Our toolkit leverages the power of oneAPI in the development of a unified interface that encapsulates the intricacies of both parallel software optimization and heterogeneous hardware optimization.

Our key contributions are:
\begin{outline}
    \1 We propose a general DRL heterogeneous training protocol that is agnostic of the types of underlying accelerators, thus portable to different heterogeneous platforms.
    \1 We develop a parameterized library that contains accelerated DRL primitives on various heterogeneous architectures (CPU, GPU, and FPGA).
    \1 We offer a Python-based User API for application developers. We also develop library interfaces for accelerator implementations. These library interfaces facilitate seamless integration of accelerated primitives with DRL application benchmark libraries at runtime.
    \1 We develop a System Composer for identifying  optimal device assignments and accelerator configurations, ensuring high performance of the DRL implementation. 
    \1 We assess our toolkit using representative DRL algorithms, DQN \cite{dqn} and DDPG \cite{ddpg}, on various benchmarks and heterogeneous platforms. 
    Compared with existing DRL frameworks, our implementations lead to a 2.1$\times$ speedup, a 3.4$\times$ system power efficiency improvement, and higher performance portability. In addition, our implementations are achieved with just dozens of lines of code, demonstrating high development productivity.

\end{outline}

\section{Background}
\setlength{\textfloatsep}{3.0pt plus 1.0pt minus 2.0pt}
\setlength{\floatsep}{3.0pt plus 1.0pt minus 2.0pt}
\setlength{\intextsep}{3.0pt plus 1.0pt minus 2.0pt}
\subsection{Deep Reinforcement Learning}
\label{sec:drlintro}
The primary time bottleneck in DRL application development is the training in simulation process \cite{meng2020efficiently}.
This process involves training a policy using software simulators before deploying it on physical agents to prevent potential physical damage during trial-and-error.
We present a generalized view of DRL training in simulation
in Figure~\ref{fig:backgnd}, 
comprising four modular components: Actors, Learner, Replay Manager (RM), and Data Storage. We refer to these modular components as DRL primitives. These primitives work and interact as follows:
\begin{figure}[h]
    \centering
    \includegraphics[width=8cm]{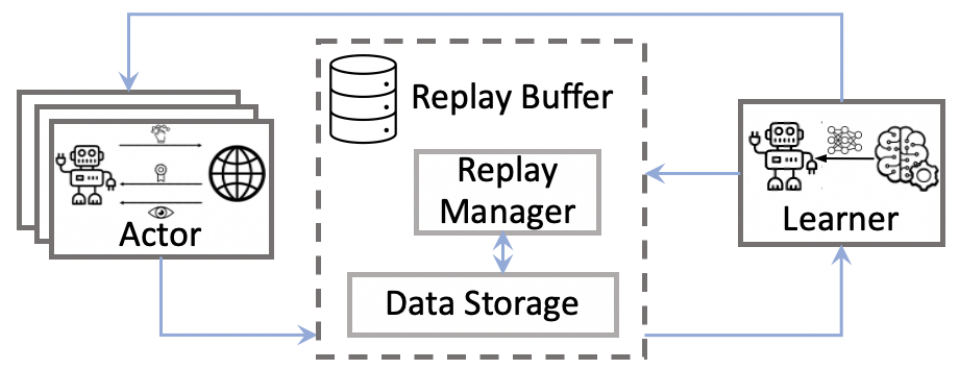}
    \caption{DRL Modular Components and Workflow}
    \label{fig:backgnd}
\end{figure}

\textbf{Actors:} Each Actor maintains a Deep Neural Network (DNN) policy network, inferring an action based on an input environment state. Each Actor operates on an individual instance of the environment simulator, applying the inferred action. The environment responds with a new state and a reward, generating a tuple \{state, action, new state, reward\}, constituting an experience (i.e., a data point) for training. Multiple copies of the Actor repeat this process to collect experiences, which populate a training dataset called the Replay Buffer.

\textbf{Replay Buffer:} The Replay Buffer serves as the dataset in DRL training. Unlike pre-labeled datasets in supervised learning, the Replay Buffer in DRL is continuously filled by online interactions of Actors with the environment, and its data points are dynamically changing as the policy evolves. 
In state-of-the-art DRL, the Prioritized Replay Buffer has gained popularity for managing data with probabilities proportional to the current policy loss to enhance training quality \cite{schaul2015prioritized, hessel2018rainbow}. It incorporates a Replay Manager (RM) associating a priority (i.e., probability of being sampled) with each experience in the Data Storage.
During data sampling, a data point (i.e., experience) $x_{i}$ is selected based on the probability distribution $\operatorname{Pr}(i)=P(i) / \sum_{i} P(i), i\in [0, \text{replay buffer size})$, where $P(i)$ represents the priority of data point $i$. This selection is achieved by identifying the minimum index $i$ for which the prefix sum of probabilities up to $i$ is greater than or equal to $x$, where $x$ is a uniformly generated random target prefix sum value between 0 and the total priority sum \cite{schaul2015prioritized}:
\begin{equation}
\label{eq:prefsum}
\scriptsize
\min _{i} \sum_{j=1}^{i} P(j) \geq x, x \sim U[0, \sum_{j=0}^{\text{replay buffer size}} P(j)]
\end{equation}
To enable rapid sampling and scalable update operations for large Data Storage, priorities are managed using a sum tree data structure \cite{zhang2023framework,  schaul2015prioritized}. Sampling and update operations on an n-ary sum tree are defined in \cite{zhang2021parallel, zhang2023framework}.

\textbf{Learner:} In each training iteration, a batch of indices are sampled via the RM to obtain data points (i.e., experiences) by reading from the Data Storage. Then, the Learner performs training using stochastic gradient descent (SGD, \cite{sgd}) on the policy network.
During the computation of the loss function in SGD,  
an updated priority is produced and written back to the Replay Buffer via the RM.
Policy network parameters are updated and sent to the Actors to ensure that experience collection employs the latest policy.

\subsection{Target Platforms}
\begin{figure}[h]
    \centering
    \includegraphics[width=6cm]{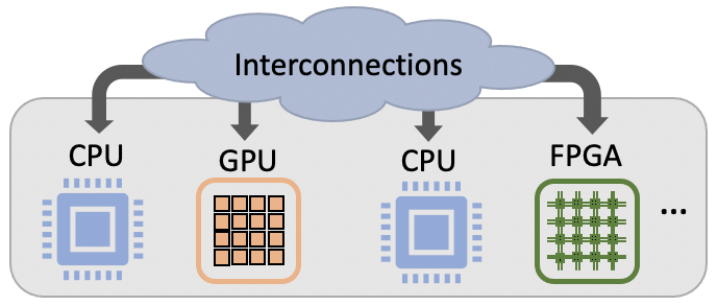}
    \caption{Heterogeneous Platform}
    \label{fig:hetarch}
\end{figure}

Today’s data centers comprise highly heterogeneous machines combining a variety of processors, accelerators, and memory \cite{hacc,devcloud,barba2021scientific}.
PEARL is designed to adapt to a wide range of heterogeneous computing platforms with interconnected CPUs and accelerators like GPUs and FPGAs, as illustrated in Figure \ref{fig:hetarch}. Developing applications on such platforms typically demands expertise in designing hardware and bridging between different programming models, which requires a learning curve that hinders the productivity of application developers. PEARL's strength lies in its ability to support DRL development across diverse heterogeneous hardware, while abstracting away complex hardware details.
\vspace{-3pt}
\subsection{Related Work}
A number of works have implemented DRL on parallel and distributed systems. RLlib introduces high-level abstractions for distributed reinforcement learning, built on top of the Ray library \cite{ray_rllib}.  
Other works, such as \cite{apex,zhang2021parallel}, implement parallel DRL algorithms by employing multiple parallel Actor threads and a centralized Learner thread, utilizing deep learning libraries like Tensorflow and LibTorch. These works leverage CPU and GPU data parallel resources for training, but do not efficiently optimize memory-bound primitives (such as small model training and replay operations) on specialized hardware.

In recent years, some research works have focused on hardware acceleration for DRL algorithms. For instance, \cite{cho2019fa3c} and \cite{meng2020accelerating} present FPGA implementations for specific algorithms, the Asynchronous Advantage Actor-Critic (A3C) and the Proximal Policy Optimization (PPO).
\cite{meng2022fpga, zhang2023framework} introduced an FPGA-based accelerator design for the Replay Buffer and mapped several DRL algorithms onto an FPGA-based heterogeneous platform.
However, these works either only optimize specific algorithms, or only target a specific heterogeneous device setup; they lack the portability to achieve high-performance implementations on different heterogeneous platforms. 
Furthermore, they lack user-friendly interfaces for DRL application developers.
Our work bridges these gaps by developing a generalized protocol that makes the development of DRL portable to different heterogeneous platforms, accompanied by a library and API that enhance productivity for DRL application developers.

\section{PEARL Toolkit}
\setlength{\textfloatsep}{2pt plus 1.0pt minus 2.0pt}
\setlength{\floatsep}{2.0pt plus 1.0pt minus 2.0pt}
\setlength{\intextsep}{2.0pt plus 1.0pt minus 2.0pt}
\begin{figure}[h]
    \centering
    \includegraphics[width=8.5cm]{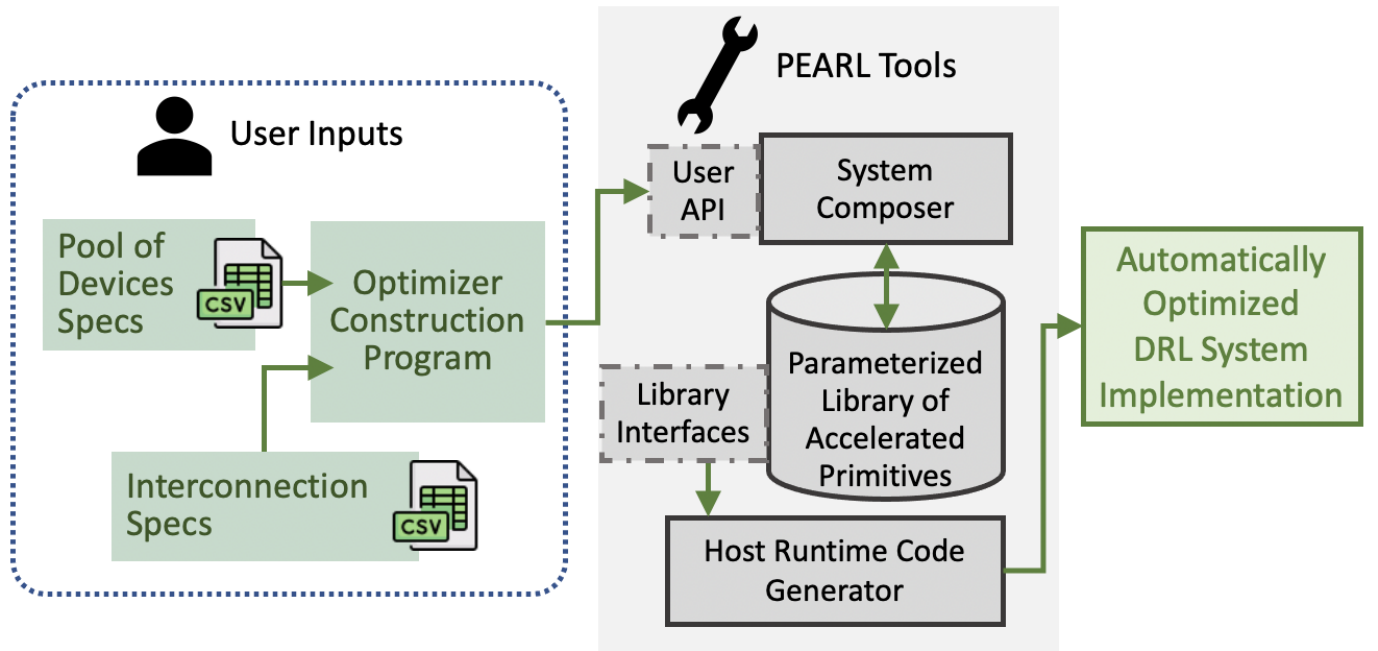}
    \caption{Toolkit Overview}
    \label{fig:toolkit}
\end{figure}
\vspace{-3pt}
\subsection{Overview}
The goal of PEARL is to provide DRL application developers with tools and familiar programming interfaces for realizing DRL training in simulation using heterogeneous platforms, while abstracting away the low-level hardware intricacies. 
As depicted in Figure \ref{fig:toolkit}, the user inputs hardware specifications and an Optimizer Construction Program; PEARL outputs a DRL system implementation that optimizes a specified metric (throughput or power efficiency). The output system implementation is a program wrapped in Python, along with any necessary accelerator executables used by the program. PEARL can be used to automatically optimize the implementations of a wide range of off-policy DRL algorithms that can be characterized by the workflow described in Section \ref{sec:drlintro} (DQN \cite{dqn}, DDPG \cite{ddpg}, SAC \cite{sac}, TD3 \cite{td3}, etc.). 
Currently, PEARL supports implementations on interconnected CPU, GPU and/or FPGA.
It provides user APIs to specify the simulation environment for Actors, the metadata of the DNN policy model, the training algorithm for the Learner, and the Replay Buffer configuration.

The PEARL toolkit implementation comprises three main components: a Host Runtime Code Generator (Section \ref{sec:runtime}), a Parameterized Library of Accelerated Primitives (Section \ref{sec:codebase}), and a System Composer (Section \ref{sec:sysconfigurator}). 
\vspace{-2pt}
\subsection{User API}
\label{sec:api}
In the Optimizer Construction Program,
PEARL provides a \texttt{Composer} object allowing the user to set input specifications, compose the system, and generate runtime code. 
Listing \ref{lst:1} shows an example Optimizer Construction Program specified by the user for optimizing the DQN algorithm \cite{dqn} on the CartPole benchmark \cite{openai_gym}:
\begin{lstlisting}[language=Python, label={lst:1},caption=Example Optimizer Construction Program]
Composer.set_device_file("/path/.csv")
Composer.set_connection_file("/path/.csv")
Composer.set_opt_metric("Throughput")
Composer.set_actors(num_actors, env="CartPole-v1")
Composer.set_train_function(learner=DQNTrainer)
Composer.set_replay(size, mode="prioritized", 
                    fanout=F)
OPT_RM, OPT_LNR = Composer.run()
Composer.RuntimeGen(replay=OPT_RM, learner=OPT_LNR)
\end{lstlisting}
All the inputs to the helper functions of the \texttt{Composer} object are hardware-agnostic, ensuring that the user inputs remain focused on algorithm and application descriptions. 

The user inputs a high-level hardware description (available devices, number of cores on the processor, amount of DSP, SRAM and Logic resources on the FPGA, bandwidth and latency of all the interconnections and memories, etc.), and a target optimization metric (throughput or power efficiency) in lines 1-3 of Listing \ref{lst:1}. 

In the \texttt{set\_actors} function, the user sets the target benchmark environment. PEARL's Actor implementations use the OpenAI Gym benchmarking library \cite{openai_gym}. Thus, providing the name of the benchmark is sufficient for using PEARL on various benchmark simulators wrapped in the Gym interface. 
In the \texttt{set\_train\_function}, the user inputs a trainer class that defines the policy model and training algorithm using Torch DNN modules \cite{pytorch}. For PEARL to parse and extract key algorithmic parameters for accelerator compilations, the input trainer class should include DNN layer definitions (e.g., using \texttt{nn.Linear} and/or \texttt{nn.Conv2d}) specifying the layer metadata of the policy model, the input tensor shape (consistent with the state shape of the Gym environment), a \texttt{get\_action} member function defining the forward pass of the DNN policy, and an \texttt{update\_policy} function for training the DNN policy using SGD \cite{sgd}. In the \texttt{set\_replay} function, PEARL provides the options of using either a Uniform Replay Buffer or a Prioritized Replay Buffer. When using a Prioritized Replay Buffer, the fanout of the sum tree in the Replay Manager needs to be specified.

By invoking the \texttt{run} function, the user obtains optimal device assignments for the Replay Manager and Learner, along with their accelerator configurations (i.e., accelerated primitive executables whose usage functions are wrapped in Python class interfaces for integration in a runtime program).
Finally, in \texttt{RuntimeGen}, the user inputs the accelerator objects produced by the \texttt{run} function to generate a complete runtime program provided as a separate output file. The resulting runtime program is ready to be employed for end-to-end DRL training in simulation on the heterogeneous platform.


\section{Training Protocol \& Runtime System}
\label{sec:runtime}
\setlength{\textfloatsep}{1.0pt plus 1.0pt minus 2.0pt}
\setlength{\floatsep}{3.0pt plus 1.0pt minus 2.0pt}
\setlength{\intextsep}{3.0pt plus 1.0pt minus 2.0pt}
\subsection{System Design}
\label{sec:runtimesys}
\begin{figure}[h]
    \centering
    \includegraphics[width=7cm]{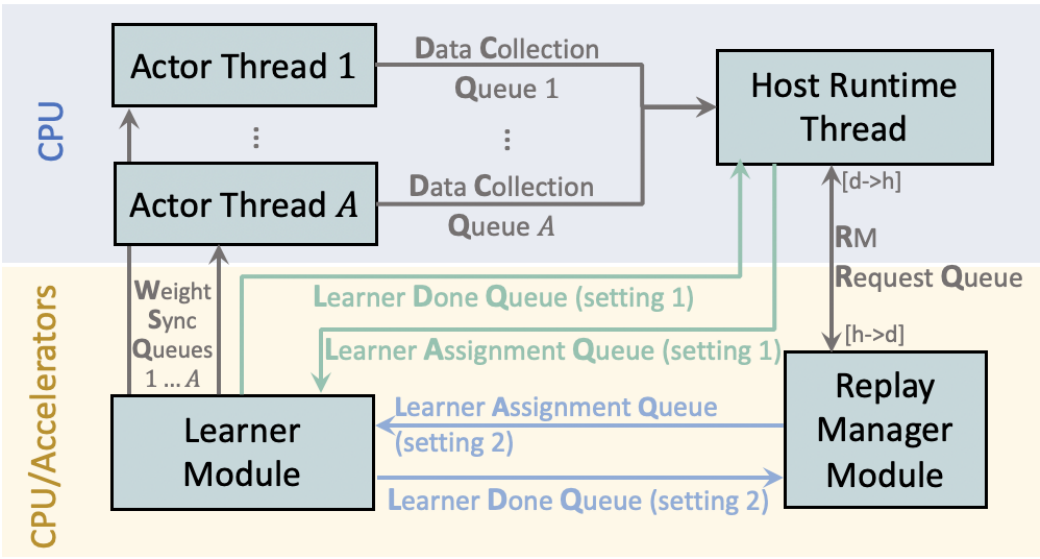}
    \caption{Runtime System}
    \label{fig:sys}
\end{figure}
\begin{figure*}[h]
    \centering
    \includegraphics[width=16cm]{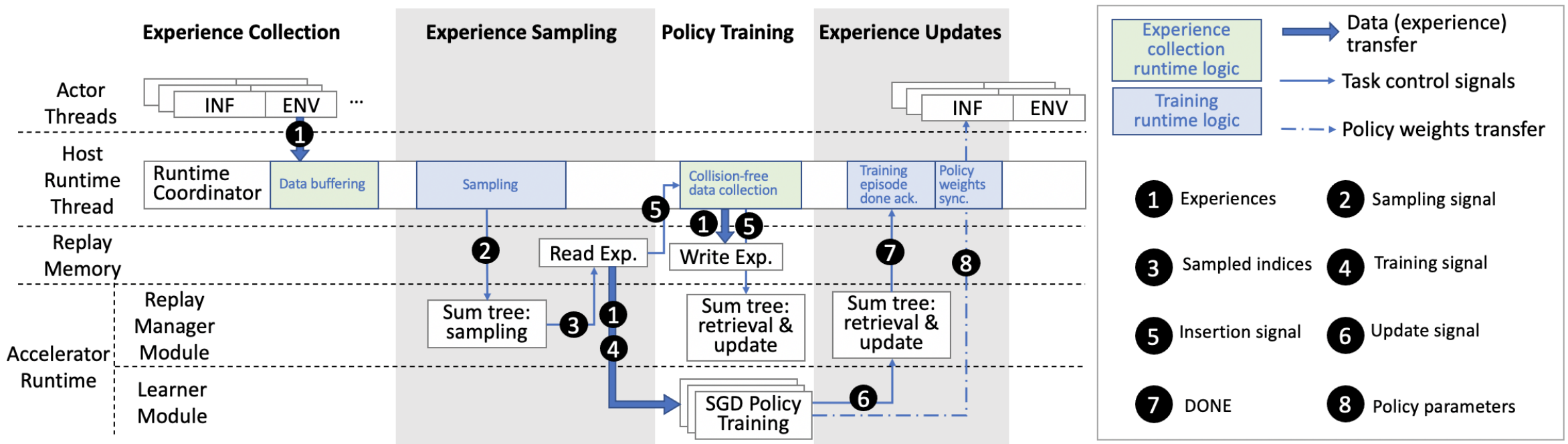}
    \caption{DRL Heterogeneous Training Protocol}
    \label{fig:protocol}
    \vspace{-1.1em}
\end{figure*}
The implementation generated by PEARL is based on a parallel DRL system managed by a Host Runtime Thread. 
Figure \ref{fig:sys} shows the setup of such a system. 
Multiple Actor threads use the policy to generate new data points (experiences) and periodically synchronize weights from the Learner. They send the experiences to the Host Runtime Thread through Data Collection Queues (DCQs). 
The Host Runtime Thread interacts with the RM through an RM Request Queue (RRQ),  where the host initiates sampling (or update) requests and receives outputs of sampled indices (or updated priorities). In an operating mode where the Learner and RM can be mapped onto arbitrary devices (i.e., setting 1), the Host Runtime Thread interacts with the Learner through a Leaner Assignment Queue (LAQ) that sends experiences and initiates training. The Learner acknowledges the completion of a training iteration using the Learner Done Queue (LDQ), and the runtime program synchronizes parameters before the next iteration. We also introduce a communication-reduction mode optimized for the cases when the RM and Learner are assigned the same device (i.e., setting 2), where the LAQ and LDQ are directly connected to the modules without communicating through the host. 
The implementations of these queues are device-dependent. Queues between CPU threads are implemented using (thread-safe) multiprocessing pipes. In our Library of Accelerated Primitives, a Queue between a CPU thread and an accelerator is implemented using \texttt{sycl::pipe} provided by oneAPI \cite{oneapi} over PCIe data transfer.  If the two modules are on the same FPGA device, the Queues between  them are implemented using on-chip FIFO pipes. If two modules are mapped to the same  GPU device, communication between them is realized by accessing a shared buffer in the global memory.
\vspace{-3pt}
\subsection{DRL Heterogeneous Training Protocol}
To perform training in simulation on a given heterogeneous system,
we propose a general DRL heterogeneous training protocol (Figure \ref{fig:protocol}).
A DRL system is implemented across three layers: the application layer consists of the DRL modular components defined in Section \ref{sec:drlintro}; the programming layer consists of libraries used to implement modular components (i.e., primitives);
the physical layer consists of the hardware and interconnection. The training protocol can be ported to various heterogeneous devices since the interactions among processors and accelerators are defined at the application layer, and are not bound to a specific type of accelerator. We show the essential data exchange and handshake signals between modular components as \circled{1}-\circled{8} in Figure  \ref{fig:protocol}. 
We provide a runtime code template that follows this protocol. It is a Python program executed on the Host Runtime Thread, and allows the ``plug and play" of heterogeneous devices for DRL primitives.

\subsection{Optimizations}
Our runtime code template also integrates a few optimizations enabled by PEARL that increase the concurrency of heterogeneous hardware resources and hide communication overheads. 
The runtime program is a while loop that iterates until a user-defined convergence criterion. 
It enables the Actor Threads, the RM Module, and the Learner Module to perform their computations in parallel by letting them continuously read from (write into) their input (output) queues,  while the runtime program handles the necessary data dependencies as it processes messages between the queues.

\subsubsection{Replay-Collision-Free Scheduling}
We adopt a strategy of deferring the immediate insertion of experiences into the Replay Buffer when experiences are received from Actor threads. We maintain a data collection buffer to cache batches of experiences, and only insert these experiences when the buffer is full. 
Upon experience insertion, we schedule the batched insertion operations after the sampling process concludes.
This optimization has two advantages. Firstly, this approach permits us to compare the insertion index against the sampled indices, hence effectively mitigating the potential contamination of data when the Learner and Actors concurrently modify the same indices of the Replay memory. We refer to this procedure as ``collision-free data collection" shown in Figure \ref{fig:protocol}. Secondly, by sequencing data insertion after the sampling phase, we align its execution concurrently with the training process. This hides the time overheads of the priority retrieval and update operations initiated by experience insertion in the training pipeline.

\subsubsection{Overlapping Computation and Communication}
As data collection (Actor threads executions) 
performs concurrently with the policy training process, the critical path in each iteration becomes the experience sampling (RM) $\rightarrow$ policy training (Learner) $\rightarrow$ experience update (RM) process. 
Other than exploiting the hardware parallelism delivered by specific devices (Section \ref{sec:codebase}), we also overlap Learner computation with replay operations and data transfers.
When using a GPU-based Learner, this is done by multi-streaming where training a sub-batch of experiences overlaps with sampling and data transfer for the next sub-batch of experiences. When using an FPGA-based Learner, this is achieved by host-device (or on-chip) streaming communication queues between the RM and the Learner, so that training using each data point starts asynchronously as soon as the Learner receives them (rather than waiting for the full batched sampling). Similarly, experience updates are also partially overlapped with the Learner.
\vspace{-3pt}
\subsection{Generated Runtime Code}
\label{sec:runtimecode}
We show an example of how the DRL training protocol is implemented in Listing \ref{lst:rt}. 
It uses the Python multiprocessing library, and launches multiple threads that exchange data, handshake, or initiate accelerator kernels.
PEARL populates data transfer and kernel launch functions using library interfaces that wraps CUDA and SYCL kernel codes in PyBind11 \cite{pybind11}. 
The library interfaces refer to the Python objects associated with the  parameterized primitives accelerators, and are used for populating the runtime program. For each primitive, the parameters exposed to the library interface are consistent with the parameters of its accelerator implementation discussed in Section \ref{sec:codebase}.
\begin{lstlisting}[language=Python, label={lst:rt}, caption=Example Generated Runtime Program Pseudo Code]
# === An Actor Thread ===
def Actor(WSQ,DCQ,env):
    while(1):
        if WSQ.poll():
            Sync_Weights(WSQ.read())
        exps = play_episode(env)
        DCQ.write(exps,insertion_requests(exps))
# === Accessing Library Interfaces ===
from FPGA_replay import PER
from GPU_learner import DQNTrainer
RM = PER(size=10000,fanout=16)
Learner = DQNTrainer(num_streams=1)
# === Host Runtime ===
# Launch Actor threads
aps = [Process(target=Actor,args=(WSQ,DCQ,env="CartPole"))]
for ap in aps: ap.start()     
# Launch Learner and RM autorun kernels
launch_learner(LAQ,LDQ,WSQ,target=Learner)
launch_RM(RRQ,target=RM)
while(1):
    # Collect data 
    for DCQ in dcqs:
        dc_buffer.append(DCQ.read())
        # Populate Replay memory (step 1) 
        if dc_buffer.full(): 
            insertion_flag = True
            break
    # After training batch completes, start priority update & sampling for next batch 
    if LDQ.poll():
        new_pr, new_weights = LDQ.read() 
        WSQ.write(new_weights)
        RRQ[h->d].write(update_requests(new_pr))
        RRQ[h->d].write(sampling_requests)
    if RRQ[d->h].poll():
        sampled_indices = RRQ[d->h].read()
        # Populate Replay memory (step 2)
        if insertion_flag:
            D_storage.CollisionFreeAdd(dc_buffer)
            RRQ.write(insertion_requests(dc_buffer))
            insertion_flag.reset()
        # Training
        sampled_data = D_storage[sampled_indices]
        LAQ.write(train_request(sampled_data))
\end{lstlisting}

\section{Parameterized Library of Primitives}
\label{sec:codebase}
\subsection{Replay Manager (RM)}
The RM performs three replay operations on a sum tree, where leaf nodes store the priorities for all experiences, and a parent node stores the sum of priorities of its children: 
\begin{itemize}
    \item Priority sampling: 
    The outputs are sampled indices (i.e., leaf nodes) based on Equation \ref{eq:prefsum}. These sampled indices are obtained by traversing the tree performing prefix sum from root to leaf. The computations are explained in \cite{zhang2021parallel}.
    \item Priority retrieval: Given the indices of the experiences, it outputs the priorities stored at the corresponding indices (i.e., leaf nodes).
    \item Priority update: the inputs are the indices of the experiences and the changes to their priorities $\Delta$; It applies the changes $\Delta$ to the priorities (and sums of priorities) stored in the corresponding nodes level-by-level from the root node down to the leaf node.
\end{itemize}
 Insertion or updating of priorities is realized with priority retrievals followed by priority updates. Drawing data from the Replay to the Learner is done by batched priority sampling. 

\subsubsection{RM on CPU and GPU}
The computations in replay operations can be viewed as a sequence of operations traversing all levels of the sum tree from the root to a leaf. 
Our RM implementations on CPU and GPU are parameterized with the tree depth, fanout, $BS$, and $W$, where $BS$ is the batch size of the replay operation requests, and $W$ is the  number of workers (degree of parallelism) allocated. Each worker is responsible for sampling or updating $\frac{B}{W}$ priorities.
All workers share concurrent accesses to the sum tree. 
We use mutex to ensure the correctness of parallel priority updates that potentially collide on the same node.
\subsubsection{RM on FPGA}
\begin{figure}[h]
    \centering
    \includegraphics[width=9cm]{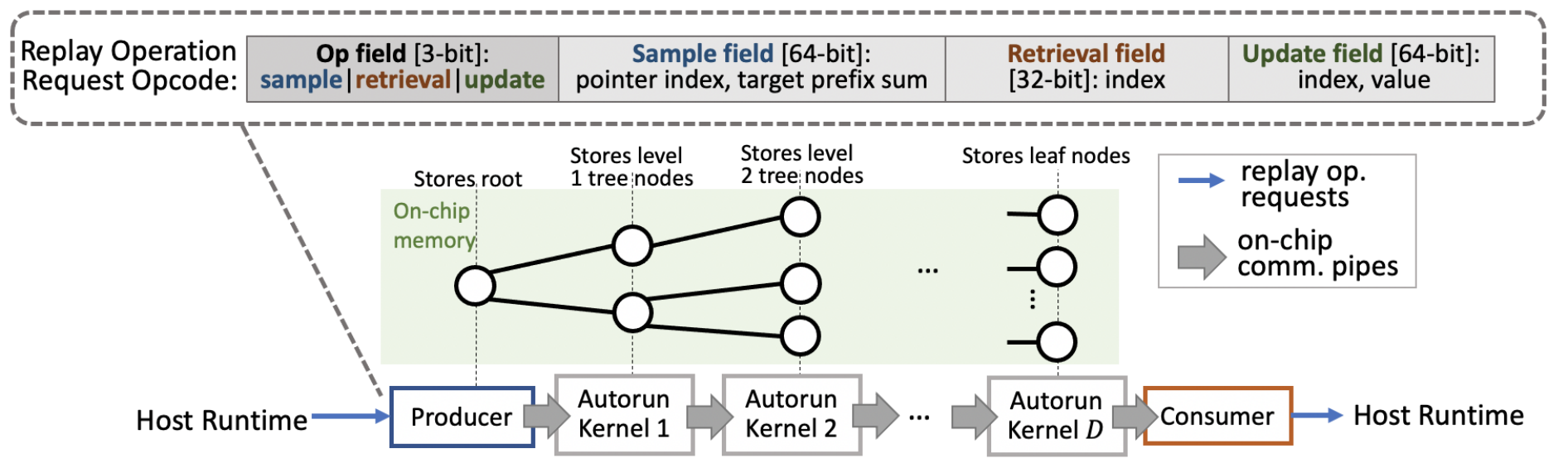}
    \caption{FPGA - Replay Manager Hardware Module}
    \label{fig:replay_fpga}
\end{figure}
We develop an accelerator template (parameterized with the tree depth and fanout) that can be re-configured to support a range of fanout and tree sizes. Typical Replay Buffer sizes range from $10K$ to $100K$, so the sum tree can be reasonably stored on-chip in block memories.
We adopt a design of multiple pipeline stages processing a stream of operation requests as shown in Figure \ref{fig:replay_fpga}. Each pipeline stage is a hardware module responsible for operating on a certain tree level and exclusively stores all the nodes on that level. 
Each hardware module is described using an Autorun kernel \cite{autorun}, which means the module operates continuously, processing data streamed to and from it at runtime. This Autorun feature eliminates the need for repeated FPGA kernel launches initiated by the host, thus avoiding this overhead in millions of DRL iterations.
Different replay operation requests in a batch are concurrently processed by different pipeline stages. The request fed into the accelerator has a unified operation code as shown in the top of Figure \ref{fig:replay_fpga}. 
The requests are decoded at each pipeline stage, and the corresponding operations are executed in an online manner.

\subsection{Learner}
The Learner takes in a batch of experiences, and performs SGD on the policy network(s). It constitutes forward propagation (FP), loss function (LOSS), backward propagation (BP), and gradient aggregation (GA). 

\subsubsection{Learner on CPU and GPU}
We use PyTorch \cite{pytorch, torchintel} to implement DNN training on CPUs and GPUs. 
On the GPU, PyTorch utilizes CuDNN \cite{pytorch} or Xe Matrix Extensions \cite{torchintel} backend to exploit SIMD parallelism. We also support using multiple streams, each stream independently processes the FP, LOSS, BP, and GA on a sub-batch of experiences. Afterward, the generated gradients from the streams are aggregated to update the weights. Compared to bulk processing a full batch of data, this helps overlap the data transfer and computation time between sub-batches of data. 
The GPU-based Learner code is parameterized to specify the number of streams.
\subsubsection{Learner on FPGA}
\begin{figure}[h]
    \centering
    \includegraphics[width=9cm]{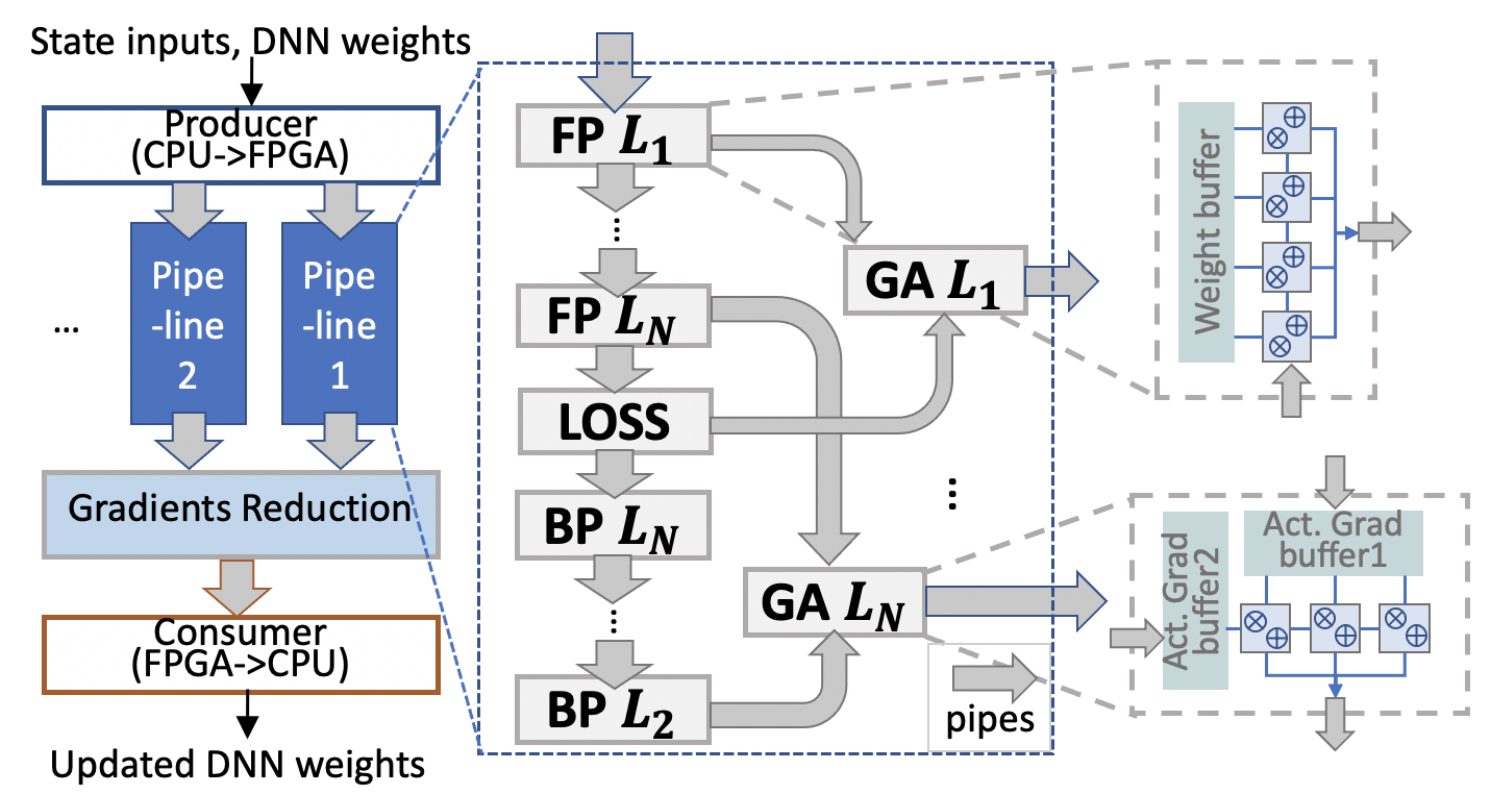}
    \caption{FPGA - Learner Hardware Module}
    \label{fig:learner_fpga}
\end{figure}
On FPGA, we design a Learner Module that supports both pipeline parallelism across different neural network layers and data parallelism among sub-batches of data. 
As an example, we show the pipeline design for an $N$-layer MLP in Figure \ref{fig:learner_fpga}. Each pipeline stage uses buffers to store intermediate activations, and uses an array of multiplier-accumulator units to compute matrix-vector multiplication for a given input. The number of multiplier-accumulator units allocated to each layer is controlled by a unique unroll factor $UF$, which will be tuned to ensure load balancing for best performance (Section \ref{sec:asetup}). These modules (denoted FP, BP, LOSS, and GA) are also described in Autorun kernels \cite{autorun} to avoid repeated kernel launch overheads in DRL loops. To realize data streaming between modules, they are connected by on-chip first-in-first-out pipes. To support data parallelism, we make $DP$ copies of such pipelines. Each pipeline generates the gradients for a sub-batch of experiences, and they are accumulated using a reduction circuit before updating the weights and sending them back to the host.
The FPGA Learner code is parameterized with the model metadata (e.g., the number of layers and the dimensions of each layer), $DP$ and a set of $UF$.
\section{System Composer}
\setlength{\textfloatsep}{1.0pt plus 1.0pt minus 1.0pt}
\setlength{\floatsep}{1.0pt plus 1.0pt minus 1.0pt}
\setlength{\intextsep}{1.0pt plus 1.0pt minus 1.0pt}
\label{sec:sysconfigurator}
Given the the user-specified Replay Manager (RM) and Learner metadata in the Optimizer Construction Program as inputs, 
the goals of the system composer are to (A) determine the best-performing accelerator configuration within each device for all the primitives, and (B) determine an optimal primitive-to-device assignment that maximizes system performance.

\subsection{Accelerator Setup and Performance Estimation}
\label{sec:asetup}
To realize goal (A), we customize the parameterized accelerators described in Section \ref{sec:codebase} to suit the user-input RM and Learner specifications. 
Customization refers to adapting the degree of parallelism assigned to the primitive on a processor or accelerator (e.g., number of compute cores allocated on CPU or GPU, number of pipeline modules and buffer sizes allocated on FPGA).
Based on the customized accelerators, we obtain the expected latency of executing each primitive in one DRL iteration on each of the available devices, and store these latency numbers in a Primitive Device Assignment Matrix for further analysis of system performance in goal (B).

\setlength{\tabcolsep}{2pt}
\renewcommand{\arraystretch}{1.5}
\begin{table}[]
\scriptsize
\centering
\caption{Primitive Device Assignment Matrix: An Example }
\label{tab:compute-perf}
\begin{tabular}{|c|c|c|c|}
\hline
\multicolumn{1}{|l|}{} & \textbf{\begin{tabular}[c]{@{}c@{}}RM only\end{tabular}} & \textbf{\begin{tabular}[c]{@{}c@{}}Learner only\end{tabular}} & \textbf{\begin{tabular}[c]{@{}c@{}}RM \& Learner\end{tabular}} \\ \hline
\textbf{CPU} & $T_{RM}(CPU)$ & $T_{Learner}(CPU)$ & $T_{RM'}$, $T_{Learner'}(CPU)$ \\ \hline
\textbf{GPU} & $T_{RM}(GPU)$ & $T_{Learner}(GPU)$ & $T_{RM'}$, $T_{Learner'}(GPU)$ \\ \hline
\textbf{FPGA} & $T_{RM}(FPGA)$ & $T_{Learner}(FPGA)$ & $T_{RM'}$, $T_{Learner'}(FPGA)$ \\ \hline
\end{tabular}
\end{table}
The Primitive Device Assignment Matrix is a 3$\times N$ table. The $N$ rows stand for the choices of $N$ available devices in the heterogeneous platform; the 3 columns correspond to the (combination of) primitive(s) to be assigned to one of the devices. An example of a Primitive Device Assignment Matrix for a single-node processor connected to two accelerators is shown in Table \ref{tab:compute-perf}. The notion of $T^x_y$ in the table entries denotes the latency of performing one iteration of the given primitive $x$ on device $y$ (For the RM, the latencies include those of the sampling, updates and insertions).  
We explain how the table entries are populated based on accelerator setups as follows:

\textbf{Primitive Setup on a CPU:} 
For the Learner, we allocate threads for training such that the time for processing $BS$ experiences matches that of generating $BS$ experiences by the Actors. 
For the RM, the degree of parallelism $W$ is ($\text{Total \# threads in the CPU - \# threads for the Learner}$), clipped by the range $[1,BS]$. Afterward, we run the primitive with synthetic data for $1K$ iterations and profile the per-iteration latency and power consumption for filling the $T(CPU)$ entries in the Primitive Device Assignment Matrix.

\textbf{Acceleration Setup on a GPU:}  
For the RM, the degree of parallelism is always set to $BS$.
The sum tree is stored in the GPU global memory.
For the Learner, we search for the best-performing number of streams in the range $[1,BS]$ and record their per-SGD-step latencies. Afterward, we use the optimal latency and power consumption of each primitive for filling the Primitive Device Assignment Matrix.

\textbf{Accelerator Configuration on an FPGA:} 
The RM and the Learner can both be mapped to the same FPGA device only if the total buffer size required by the RM and Learner modules is smaller than the total amount of SRAM resources. This is to avoid efficiency losses in accesses to off-chip memory. For the RM, the number of Autorun kernels in the pipeline is configured to match the tree depth, and the buffer sizes are configured based on their corresponding tree levels. For the Learner, the number of pipelines $DP$ is set to the largest value such that 
\vspace{-3pt}
\begin{multline}
\small
    DP\times\text{total buffer size of one pipeline}\leq\\
\small\text{total SRAM size}-\text{SRAM size consumed by the RM}.
\end{multline}
\vspace{-3pt} 
The amount of compute resources allocated to each pipeline stage, $UF$, is tuned such that all pipeline stages are load balanced (for the maximal effective hardware utilization): 
\begin{multline}
\small
    T_{stage}=\frac{\# MAC^{\text{FP L}_1}}{UF^{\text{FP L}_1}} =
    ... =\frac{\# MAC^{\text{GA L}_N}}{UF^{\text{FP L}_N}};\\
\small
    \text{where }UF^{\text{FP L}_1}+UF^{\text{FP L}_2}+...+UF^{\text{FP L}_N}\leq \frac{\text{\#DSPs}}{DP}.
\end{multline}
We obtain the latency of accelerators on FPGA through performance modeling:
\begin{gather}
    T_{RM}^\text{sampling}=2\times F\times (BS+D)+T_{comm}^{(i\rightarrow FPGA)}\\
    T_{RM}^\text{update or insert}=2\times (BS+D)+T_{comm}^{(i\rightarrow FPGA)}\\
    T_{Learner}=T_{stage}\times (BS+3\times(\text{\#layers}-1))
\end{gather}
In equations 4-6, the pipeline latencies are calculated using the following steps: First, we determine the per-stage latency. Next, we multiply this per-stage latency by the sum of batch size $BS$ and pipeline overhead $D$, where $D$ is the pipeline fill/drain overhead ($D$ is directly proportional to the depth of the pipeline). In the Replay Manager (RM), $D$ equals the sum tree depth, while in the Learner, it equals the total number of layer propagations.
$T_{comm}$ refers to the communication time of taking inputs from other primitives residing on the same or different device, which can be modeled as $lat+\frac{\text{data size}}{\text{bandwidth}}$. The data transfer latency overhead ($lat$) and bandwidth characteristics will be filled later in Algorithm \ref{algo:mapping} - Equation \ref{eq:obj} depending on whether the communication is within the same device (e.g., through DDR) or across different devices (e.g., through PCIe). In Equation 4, $i$ refers to the device that initiates sampling. In Equation 5, $i$ refers to either the Actors device (CPU) for insertion requests or the Learner device for update requests.

\vspace{-2pt}
\subsection{Heterogeneous System Composition Algorithm}
\label{sec:mapping}

\begin{algorithm}
\small
    \caption{Heterogeneous System Composition Algorithm}
    \label{algo:mapping}
    \begin{algorithmic}[1]
        \STATE {\bfseries Input:} Primitive Device Assignment Matrix $M$, 
        \STATE \texttt{\# Step 1: Primitive Placement}
         \IF{Optimizing throughput}
        \STATE D[{\text{RM}}], $D[{\text{Learner}}]=argmax_{i,j}\{\frac{\text{Iteration Batch Size}}{T_\text{itr}}\}$
        \ELSIF{Optimizing power efficiency}
        \STATE D[{\text{RM}}], $D[{\text{Learner}}]=argmax_{i,j}\{\frac{\text{Iteration Batch Size}}{T_\text{itr}\times\sum \text{power}}\}$
         \ENDIF
        \STATE where $i,j$ denotes available devices for RM and Learner in $M$,
        \begin{equation}
        \small
        \label{eq:obj}
        T_{\text {itr}}=\eqnmark{comm1}{T_{RM}^{\text {sampling}}(i)}+\max (\eqnmark{comm3}{T_{R M}^{\text {insert}}(i)}, \eqnmark{comm2}{T_{RM}^{\text{update}}(i)}+T_{\text {Learner}}(j))
        \end{equation}
\annotate[yshift=0em]{below,right}{comm1}{$T_{\text {comm}}^{(i \rightarrow j)}$}
\annotate[yshift=0em]{below,right}{comm2}{$T_{\text{comm}}^{(j \rightarrow i)}$}
\annotate[yshift=0em]{below,right}{comm3}{$T_{\text {comm }}^{(cpu \rightarrow i)}$}
        \STATE \textbf{Output} $D[\text{Learner}], D[\text{RMM}]$
        \STATE \texttt{\# Step 2: Memory Component Placement}
        \STATE Initialize $D[\text{Data Storage}]$; min$\_$traffic$\leftarrow \infty$
        \STATE $C_{\text{Learner}}\leftarrow B\times (E+1)$; $C_{\text{Actor}}\leftarrow N_{Actor}\times E$; $C_{\text{RM}}\leftarrow B$
        \FOR{$i$ in [Learner, Actors, RM]} 
         \STATE Total data traffic = $\sum^{i'\in\{\text{Learner,Actors,RM}\}} \frac{C_{i'}}{\text{bandwidth}(D[i],D[i'])}$
         \IF{Total data traffic $<$ min$\_$traffic}
         \STATE min$\_$traffic $\leftarrow$ Total data traffic;  $D[\text{Data Storage}]\leftarrow D[i]$
         \ENDIF
        \ENDFOR
        \STATE \textbf{Output} $D[\text{Data Storage}]$
        
    \end{algorithmic}
\end{algorithm}
Based on a completed Primitive Device Assignment Matrix, we develop a Heterogeneous System Composition Algorithm (Algorithm \ref{algo:mapping}) to choose an optimal combination of devices for assigning each primitive.
It first determines the best device assignment of the primitives to maximize the target metric,
then places the memory component (Data Storage) to minimize the total data traffic in the system.

In Step 1 (lines 2-9, Algorithm \ref{algo:mapping}), the training throughput can be estimated using the processed batch size in each iteration, $BS$, and the iteration execution time, $T_{itr}$. $T_{itr}$ is defined in Equation \ref{eq:obj}. 
The critical path in an iteration is the priority sampling followed by SGD training and priority update, while the other replay operations overlap with the training process. 
The required costs of communication with other compute modules are encapsulated in each component of Equation \ref{eq:obj} corresponding to the candidate devices $i,j$ for RM and Learner, where $i,j$ are permutated to include all the device assignment choices. When $i=j$, the latencies are sampled from the third column of the Compute-Performance Table (e.g., Table \ref{tab:compute-perf}). The complexity of Step 1 is $O(M^2)$, where $M$ is the number of available devices on the heterogeneous platform.

In Step 2 (lines 10-19, Algorithm \ref{algo:mapping}), we decide on the device assignment of the Data Storage. The data traffic wrt the Data Storage during each iteration includes $BS$ words of sampling indices from the $D_\text{Learner}$, $BS\times E$ sampled experiences to the $D_\text{Learner}$ (where $E$ is the size of each experience for the given benchmark), and $N_{Actor}\times E$ inserted experiences from the Actors. These communication costs are denoted as $C$ in Algorithm \ref{algo:mapping}. We place Data Storage on the device that minimizes the total data traffic based on available bandwidths between devices (e.g., PCIe) and within each device (e.g., DDR). The complexity of Step 2 is $O(1)$, as the number of primitives in a DRL algorithm is constant.

\section{Evaluation}
\setlength{\textfloatsep}{2.0pt plus 1.0pt minus 0.0pt}
\setlength{\floatsep}{3.0pt plus 1.0pt minus 0.0pt}
\setlength{\intextsep}{2.0pt plus 1.0pt minus 1.0pt}
\vspace{-2pt}
\subsection{Experiment Setup}
\paragraph*{\textbf{Hardware Platforms}}
To show the portability of our toolkit to different platforms, we conduct our experiments on two heterogeneous platforms. The first platform, $Server_{CG}$, has a Host CPU  and an integrated GPU that shares the same die. The second platform, $Server_{CGF}$, consists of a Host CPU connected to a GPU and an FPGA, both through PCIe with 16 GB/s bandwidth. The specifications of these platforms are summarized in Table \ref{tab:devices}. 
For FPGA bitstream generation, we follow the oneAPI development flow \cite{oneapi}.
\setlength{\tabcolsep}{4pt}
\renewcommand{\arraystretch}{1.1}
\begin{table}[]
\centering
\scriptsize
\caption{Specification of Heterogeneous Platforms}
\label{tab:devices}
\setlength\extrarowheight{-1pt}
\begin{tabular}{|c|cc|ccc|}
\hline
\textbf{Platform} & \multicolumn{2}{c|}{\textbf{$Server_{CG}$}} & \multicolumn{3}{c|}{\textbf{$Server_{CGF}$}} \\ \hline
\textbf{Device} & \multicolumn{1}{c|}{\textbf{\begin{tabular}[c]{@{}c@{}}CPU\\ Intel Core \\ i9-\\ 11900KB\end{tabular}}} & \textbf{\begin{tabular}[c]{@{}c@{}}GPU\\ Intel UHD \\ Graphics \\ Xe\end{tabular}} & \multicolumn{1}{c|}{\textbf{\begin{tabular}[c]{@{}c@{}}CPU\\ Intel\\ Xeon \\ Gold 6326\end{tabular}}} & \multicolumn{1}{c|}{\textbf{\begin{tabular}[c]{@{}c@{}}GPU\\ Nvidia\\ Geforce\\ 3090\end{tabular}}} & \textbf{\begin{tabular}[c]{@{}c@{}}FPGA\\ Intel \\ DE10-\\ Agilex\end{tabular}} \\ \hline
\textbf{Processs} & \multicolumn{1}{c|}{10 nm} & 10 nm & \multicolumn{1}{c|}{10 nm} & \multicolumn{1}{c|}{8 nm} & 10 nm \\ \hline
\textbf{\begin{tabular}[c]{@{}c@{}}Hardware\\ Parallelism\end{tabular}} & \multicolumn{1}{c|}{\begin{tabular}[c]{@{}c@{}}2 sockets,\\ 16 cores\end{tabular}} & \begin{tabular}[c]{@{}c@{}}32 Unified\\ Pipelines\end{tabular} & \multicolumn{1}{c|}{\begin{tabular}[c]{@{}c@{}}2 sockets,\\ 64 cores\end{tabular}} & \multicolumn{1}{c|}{\begin{tabular}[c]{@{}c@{}}10496 \\ CUDA Cores\end{tabular}} & \begin{tabular}[c]{@{}c@{}}4510 \\ DSPs\end{tabular} \\ \hline
\textbf{\begin{tabular}[c]{@{}c@{}}External \\ Memory\end{tabular}} & \multicolumn{1}{c|}{32 GB} & 32 GB & \multicolumn{1}{c|}{\begin{tabular}[c]{@{}c@{}}1 TB,\\ DDR4\end{tabular}} & \multicolumn{1}{c|}{\begin{tabular}[c]{@{}c@{}}24 GB,\\ HBM\end{tabular}} & \begin{tabular}[c]{@{}c@{}}32 GB, \\ DDR4\end{tabular} \\ \hline
\textbf{Frequency} & \multicolumn{1}{c|}{3.3 GHz} & 1.6 GHz & \multicolumn{1}{c|}{2.9 GHz} & \multicolumn{1}{c|}{1.7 GHz} & 400 MHz \\ \hline
\end{tabular}
\end{table}
\begin{table}[h]
\centering
\caption{Benchmarking Environments and Algorithms}
\label{tab:nn_workload}
\setlength\extrarowheight{-1pt}
\begin{tabular}{|c|cccc|}
\hline
\textbf{Environment} & \textbf{Algorithm} & \textbf{\begin{tabular}[c]{@{}c@{}}State\\ Dim.\end{tabular}} & \textbf{\begin{tabular}[c]{@{}c@{}}Action\\ Dim.\end{tabular}} & \textbf{\begin{tabular}[c]{@{}c@{}}DNN\\ Policy\end{tabular}} \\ \hline
\textbf{CartPole} & DQN & $4$ & $1$ & \begin{tabular}[c]{@{}c@{}}3-layer MLP, \\ hidden size 64\end{tabular} \\ \hline
\textbf{MountainCar} & DDPG & $8$ & $4$ & \begin{tabular}[c]{@{}c@{}}4-layer MLP,\\ hidden sizes 256,128\end{tabular} \\ \hline
\textbf{Pong} & DQN & $84\times84$ & $6$ & CNN in \cite{dqn} \\ \hline
\end{tabular}
\end{table}
\paragraph*{\textbf{Algorithms and Benchmarking Software Environments}}
We select three widely-used RL benchmarking environments: discrete-action classic control task CartPole, continuous-action task MountainCar, and Atari games Pong, 
in the OpenAI Gym software simulation environment \cite{openai_gym}.
We demonstrate our toolkit using two representative DRL algorithms widely applied in various applications, DQN \cite{dqn} and DDPG \cite{ddpg}. 
The algorithm, size of the states and actions, and policy model for solving each benchmark are shown in Table~\ref{tab:nn_workload}.  
Note that on $Server_{CG}$, we only tested for the CartPole and MountainCar benchmarks due to the current lack of library support for CNN training on integrated GPUs.

\paragraph*{\textbf{Performance Metrics}}
We evaluate the two optimization metrics supported by PEARL: (1) Training throughput is the number of Experiences processed Per Second ($EPS=\frac{\text{Training batch size}}{T_{itr}}$, where $T_{itr}$ is the execution time of one training iteration defined in Equation \ref{eq:obj}); (2) Power efficiency ($EPS/Watt$) is computed by dividing $EPS$ with the total power consumption of the heterogeneous devices, memory, and interconnections used.

\subsection{Performance of Accelerated Primitives}
Since the throughput $EPS$ is bounded by latencies of the primitives in each DRL training iteration, we first show the device assignment tradeoffs for each primitive.

\begin{figure}[h]
\captionsetup[subfigure]{aboveskip=-3pt,belowskip=1pt}
    \centering
    \begin{subfigure}{8.5cm}
        \includegraphics[width=\linewidth]{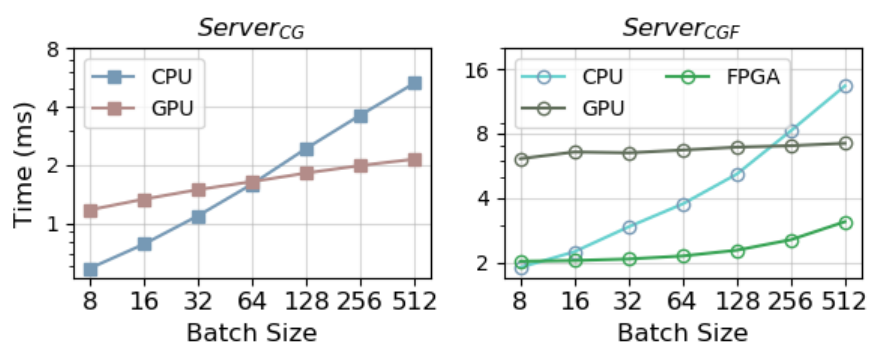}
        \caption{Priority Sampling}
        \label{fig:sample_exp}
    \end{subfigure}
    \begin{subfigure}{8.5cm}
        \includegraphics[width=\linewidth]{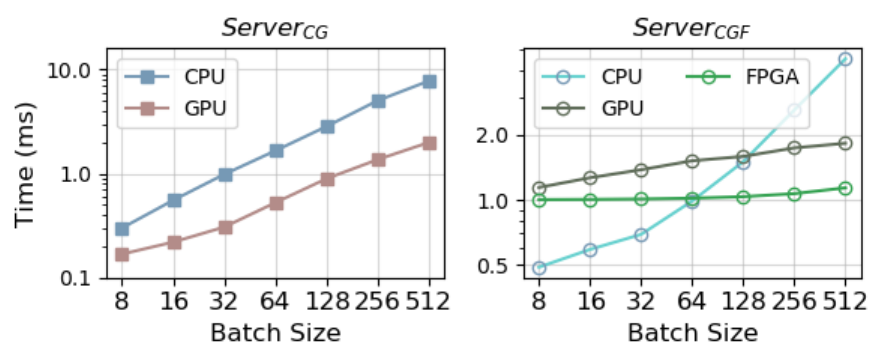}
        \caption{Priority Retrieval and Update}
        \label{fig:update_exp}
    \end{subfigure}
    \caption{Replay Manager}
    \label{fig:combined_exp}
\end{figure}

In Figures \ref{fig:combined_exp}, we present the total execution latencies for batched Replay Manager (RM) operations. They are plotted across a range of commonly-used training batch sizes (a significant DRL hyper-parameter affecting DRL iteration time).
For PCIe-connected GPU and FPGA on $Server_{CGF}$, all the latencies of primitives in Figure \ref{fig:combined_exp} include the data transfer (PCIe) time.
Note that the latencies for priority retrieval and update are combined since these operations are typically performed together during priority insertion or update processes. 
Our observations reveal superior scalability of GPU- and FPGA-accelerated replay operations compared to the multi-threaded CPU implementation. The RM operations are memory-bound, with Priority sampling having an arithmetic intensity of $F$ FLOPS/word and Priority update operating at $1$ FLOP/word (where $F$ represents the RM sum tree fanout). 
While GPU data parallel compute resources exhibit good scalability, they are underutilized due to high-latency global memory accesses that cannot be hidden by the computations. In contrast, the FPGA accelerator processes the sum tree operations in a near-memory manner, storing the data structure on-chip, thus delivering the highest scalability.

\begin{figure}[h]
    \centering
    \includegraphics[width=9cm]{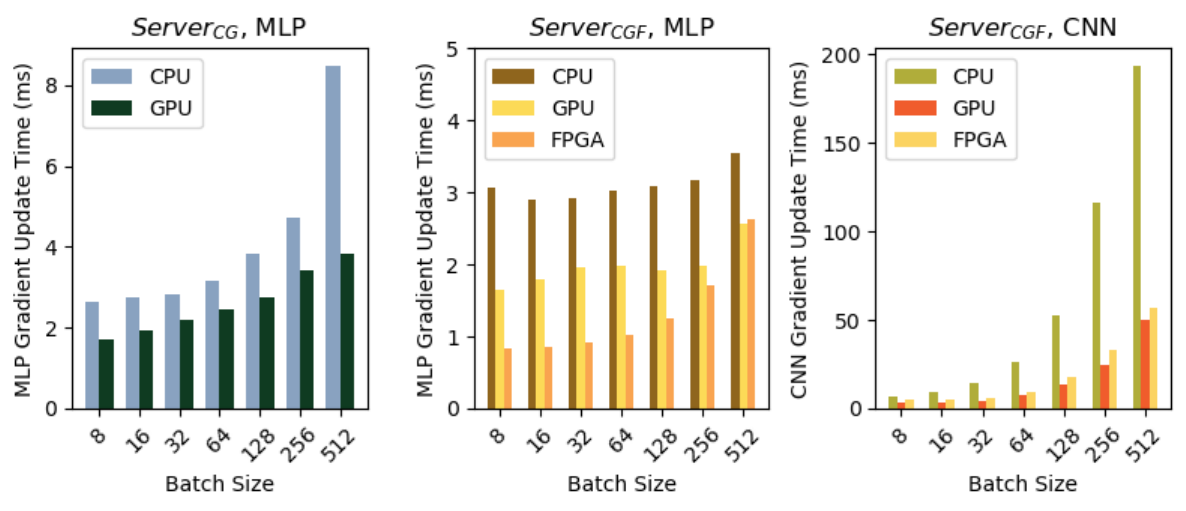}
    \caption{Learner}
    \label{fig:learner_exp}
\end{figure}
\begin{figure*}
\captionsetup[subfigure]{aboveskip=-3pt,belowskip=1pt}
    \centering
    \begin{subfigure}{0.24\textwidth}
        \includegraphics[width=\linewidth]{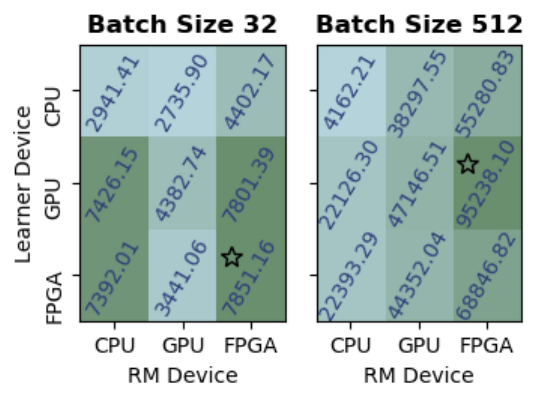}
        \caption{DDPG - MC, $Server_{CGF}$}\label{fig:ll_thp}
    \end{subfigure}%
    \hfill
    \begin{subfigure}{0.24\textwidth}
        \includegraphics[width=\linewidth]{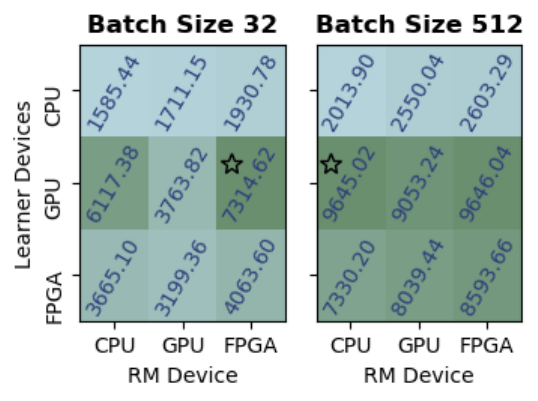}
        \caption{DQN - Pong, $Server_{CGF}$}\label{fig:pong_thp}
    \end{subfigure}
    \hfill
    \begin{subfigure}{0.22\textwidth}
        \includegraphics[width=\linewidth]{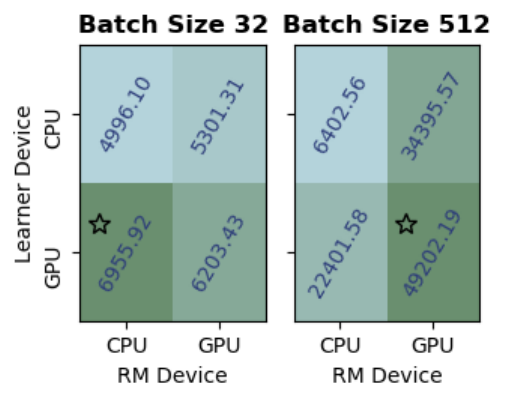}
        \caption{DQN - CP, $Server_{CG}$}\label{fig:devc_cp_thp}
    \end{subfigure}
    \hfill
    \begin{subfigure}{0.22\textwidth}
        \includegraphics[width=\linewidth]{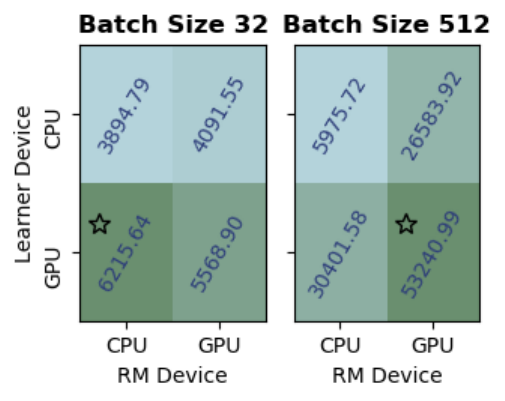}
        \caption{DDPG - MC, $Server_{CG}$}\label{fig:devc_ll_thp}
    \end{subfigure}
    
    \begin{subfigure}{0.24\textwidth}
        \includegraphics[width=\linewidth]{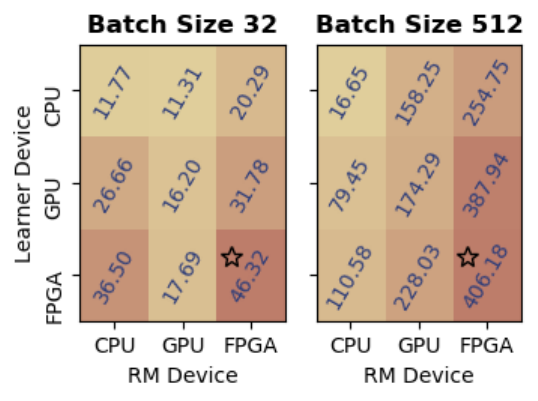}
        \caption{DDPG - MC, $Server_{CGF}$}\label{fig:ll_pe}
    \end{subfigure}%
    \hfill
    \begin{subfigure}{0.24\textwidth}
        \includegraphics[width=\linewidth]{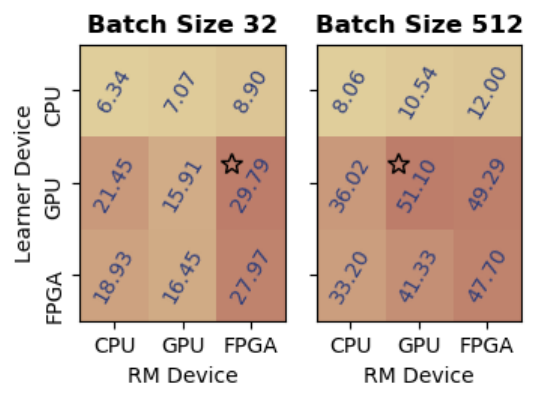}
        \caption{DQN - Pong, $Server_{CGF}$}\label{fig:pong_pe}
    \end{subfigure}%
    \hfill
    \begin{subfigure}{0.22\textwidth}
        \includegraphics[width=\linewidth]{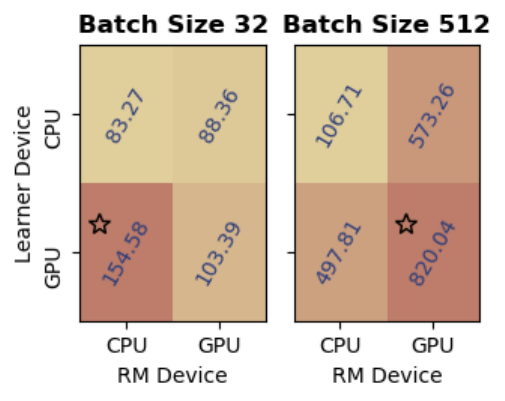}
        \caption{DQN - CP, $Server_{CG}$}\label{fig:devc_cp_pe}
    \end{subfigure}
    \hfill
    \begin{subfigure}{0.22\textwidth}
        \includegraphics[width=\linewidth]{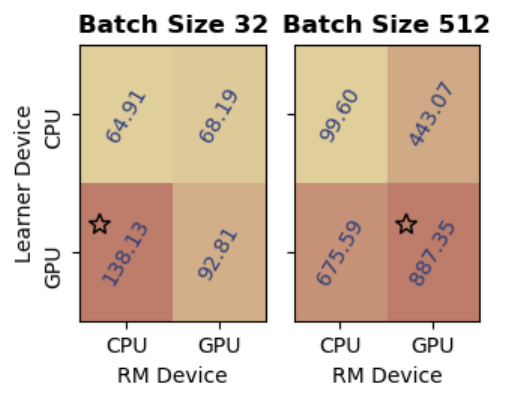}
        \caption{DDPG - MC, $Server_{CG}$}\label{fig:devc_ll_pe}
    \end{subfigure}
    \caption{System Composition. (a)-(d): Throughput ($EPS$). (e)-(h): Power Efficiency ($EPS/Watt$). }
    \label{fig:syscomposer}
    \vspace{-0.5em}
\end{figure*}
In Figure \ref{fig:learner_exp}, we show the Learner execution times for one gradient update iteration.
Batched layer propagations involve dense matrix multiplications, which exhibit a higher arithmetic intensity compared to replay operations. Consequently, the advantages of utilizing data parallel architectures (GPUs) are evident, with consistently lower gradient update latency compared to CPU execution.
The FPGA accelerator design surpasses GPU performance when arithmetic intensity is low. 
This is particularly evident when dealing with smaller neural network sizes and batch sizes.
The GPU incurs notable kernel and memory overhead that cannot be hidden during low arithmetic intensity operations.
Nonetheless, as the batch size increases, the execution time of training primitives on GPU begins to outperform that on FPGA. This shift is due to hidden memory overhead at larger batch computations and a higher clock frequency on the GPU.

\begin{table*}[h]
\caption{Comparison with Existing DRL Frameworks}
\label{tab:existing}
\scriptsize
\centering
\begin{tabular}{lc|c|c|c|c|c|c|c|c|c}
 &                                                                             & \multicolumn{3}{c|}{\textbf{DQN-CartPole}}                                                & \multicolumn{3}{c|}{\textbf{DDPG-MountainCar}}                                            & \multicolumn{3}{c}{\textbf{DQN-Pong}}                                                   \\ \cline{3-11} 
 &                                                                             & PEARL         & RLlib        & \begin{tabular}[c]{@{}c@{}}Stable\\ Baselines3\end{tabular} & PEARL         & RLlib        & \begin{tabular}[c]{@{}c@{}}Stable\\ Baselines3\end{tabular} & PEARL        & RLlib       & \begin{tabular}[c]{@{}c@{}}Stable\\ Baselines3\end{tabular} \\ \cline{2-11} 
 & \textbf{\begin{tabular}[c]{@{}c@{}}$EPS$\\ Batch 32, 512\end{tabular}}      & 7.8K, 90.3K & 4.1K, 50.3K & 4.6K, 58.1K                                                & 7.8K, 95.2K & 3.6K, 51.5K & 4.3K, 60.1K                                                & 7.3K, 9.6K & 7.0K, 7.2K & 5.2K, 6.9K                                                 \\ \cline{2-11} 
 & \textbf{\begin{tabular}[c]{@{}c@{}}$EPS/Watt$\\ Batch 32, 512\end{tabular}} & 46.1, 473.8 & 14.2, 189.9 & 16.5, 219.3                                                & 46.3, 406.2 & 13.6, 194.3 & 17.24, 227.0                                               & 29.8, 51.1 & 19.1, 24.3 & 17.2, 22.8                                                 \\ \cline{2-11} 
 & \textbf{\begin{tabular}[c]{@{}c@{}}$\Phi(D)$\\ Batch 32, 512\end{tabular}} & 4.9K, 21.4K & 0, 0 & 0, 0                                        
 & 4.5K, 20.3K & 0, 0 & 0, 0                                             
 & 2.9K, 4.6K & 0, 0& 0, 0   \\ \cline{2-11} 
  & \textbf{\begin{tabular}[c]{@{}c@{}}$\Phi(P)$\\ Batch 32, 512\end{tabular}} & 7.3K, 63.7K & 3.7K, 55.8K & 3.5K, 50.2K                                     
 & 6.9K, 68.2K & 2.88K, 46.8K & 3.3K, 44.5K                                     
 & 6.1K, 8.9K & 6.4K, 6.8K& 4.0K, 8.1K
\end{tabular}
\vspace{-1.3em}
\end{table*}
\subsection{System Composition}
In Figures \ref{fig:syscomposer}, we show the achieved throughput and power efficiency respectively for all device assignment choices, as well as the compositions returned by the PEARL toolkit, on both heterogeneous platforms. 
In all the subfigures, the color gradients in the grids are proportional to the magnitudes of achieved throughput in their corresponding device assignments.
The throughput is highly dependent on the RM and Learner iteration latencies. 
We observe that the choice of device for the primitive with the highest latency significantly influences variations in throughput.
Specifically, for small-batch computations (i.e., grid plots with batch size 32),
the color gradient changes most drastically along the horizontal axis, because replay operations result in significant overheads as Learner computations are small; On the other hand, for large-batch computations (i.e., grids with batch size 512),
the color gradient changes most drastically along the vertical axis, as the Learner dominates each training iteration and replay operation overheads are hidden. Note that when multiple device assignment choices lead to the same throughput, our toolkit selects the one with the lowest total data traffic (e.g., Figure \ref{fig:pong_thp}). 
When focusing on optimizing power efficiency, the assignment of FPGA-based accelerators stands out as the preferred choice for scenarios dominated by memory-intensive operations (e.g., small models and batch sizes) in Figures \ref{fig:ll_pe} and \ref{fig:pong_pe}. The selection of GPU becomes favorable when the performance gain in training large models outweighs the power increase compared to using FPGAs in the system.
\subsection{Comparison with Existing DRL Libraries}
We compare PEARL-generated optimal implementations with two state-of-the-art DRL frameworks, RLlib \cite{ray_rllib} and OpenAI Stable Baselines 3 (SB3) \cite{stable-baselines3}, on the same set of heterogeneous hardware, $Server_{CGF}$. The performance of RLlib and SB3 are obtained using the optimal settings required by each of them (i.e., using GPU for training). The detailed performance across different benchmarks are shown in Table \ref{tab:existing}. 
\paragraph*{\textbf{System Throughput \& Power Efficiency}}
As none of the existing frameworks support FPGA-accelerated Replay or Learner operations, this additional flexibility of using hardware-optimized accelerators enables PEARL to achieve up to 1.9$\times$, 2.1$\times$ and 1.4$\times$ improvements in terms of $EPS$ for the three benchmarks. Additionally, we achieve up to 3.3$\times$, 3.4$\times$ and 2.2$\times$ improvements in terms of power efficiency for these benchmarks.
Another study focused on mapping DRL onto FPGA-based heterogeneous platforms \cite{zhang2023framework}, and evaluated using the CartPole benchmark. 
Due to the different hardware used and different optimal device assignments, the throughput is not directly comparable between our work and \cite{zhang2023framework}. Nonetheless, we compare the effective heterogeneous resource utilization (achieved throughput given the peak throughput of all the processors and accelerators in the target platform). For CartPole DQN training with batch size 32, PEARL achieves 7.8K $EPS$ using a CPU and an FPGA with a total peak performance of 0.46 TFLOPS; \cite{zhang2023framework} achieved an amortized throughput of 7.2K $EPS$ using a CPU and an FPGA with 0.72 TFLOPS. 
Despite having 36\% lower available peak device performance, our result shows a 1.72$\times$ higher $EPS$. This indicates that our task scheduling can better saturate the given heterogeneous environment compared to \cite{zhang2023framework}.
\paragraph*{\textbf{Portability}} To show the performance portability of our toolkit, we adopt the portability metric for a framework to be consistent with that described in \cite{pennycook2021navigating}:
\vspace{-4.5pt}
\begin{equation}
\Phi(H)=\left\{\begin{array}{ll}
0 & \text { if, } \exists i\in D, EPS_i=0\\
\frac{|H|}{\sum_{i \in H} \frac{1}{EPS_i}} & \text { otherwise }
\end{array}\right.
\vspace{-4.5pt}
\end{equation}
where $H$ can be either $D$ or $P$: $D$ denotes a set of device assignment choices in using a single heterogeneous platform; $P$ denotes a set of heterogeneous platforms;
$EPS_i$ is the achieved $EPS$ using the $i^{th}$ device assignment choice or platform in the set $H$. If the implementation cannot be portable to the $i^{th}$ device assignment choice or platform in the set $H$,  $EPS_i=0$. 
The results are shown in the last two rows of Table \ref{tab:existing}. 
$\Phi(D)$ quantizes the ability to use different heterogeneous resources given by a single platform. Other existing works that do not support accelerated RM or FPGA-based Learner are not portable to these device assignments ($\exists i\in D, EPS_i=0 $), thus having $\Phi(D)=0$. In contrast, our work is portable to all assignment choices provided by $Server_{CGF}$. 
Our work leverages oneAPI in the accelerated primitive development, which enables the ability to utilize compute powers of a wider range of heterogeneous devices, thus achieving better device portability and higher performance.
$\Phi(P)$ quantizes the ability to achieve high performance across different heterogeneous platforms (i.e., both $Server_{CG}$ and $Server_{CGF}$), where $EPS_i$ is the highest throughput achieved on the $i^{th}$ platform. Our toolkit consistently achieves higher platform-throughput portability $\Phi(P)$ compared with both the existing works.
\begin{figure}[h]
    \centering
    \includegraphics[width=0.8\linewidth]{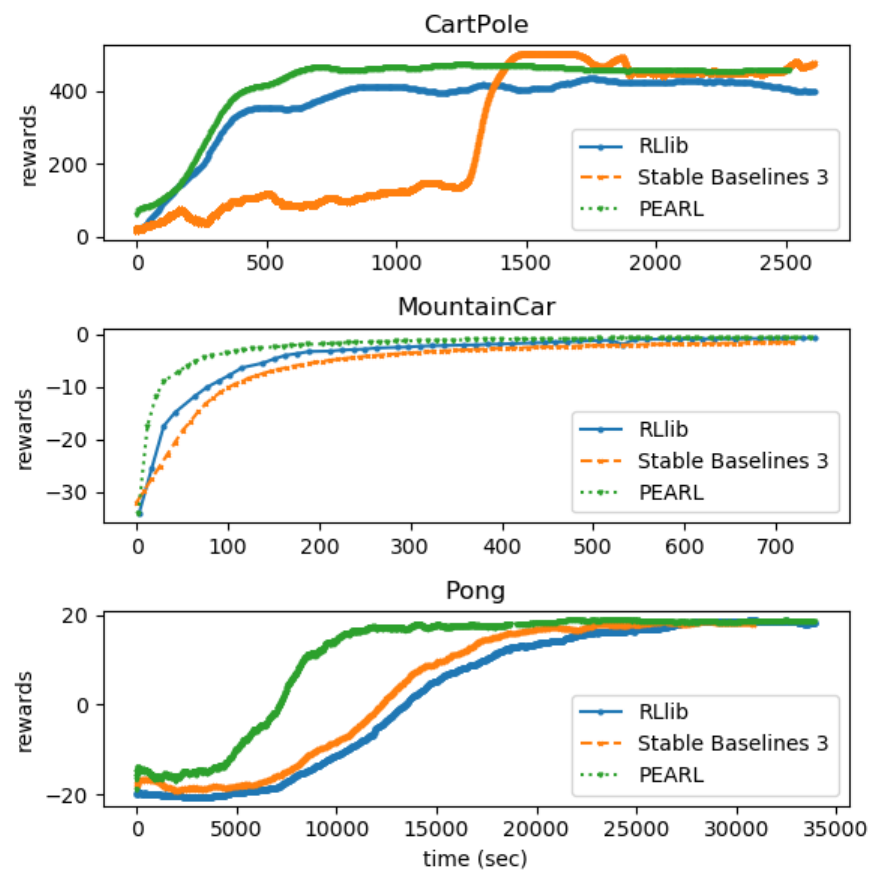}
    \caption{Rewards over Time}
    \label{fig:rewards}
\end{figure}
\paragraph*{\textbf{Algorithm Performance}}
Figure \ref{fig:rewards} plots the cumulative rewards collected by the agent policy over wall clock time.
The curves are smoothed to show the sliding average rewards obtained in a window of 100 training iterations, and each curve is the mean of 5 runs of the algorithm-benchmark pair. For all the algorithms and benchmark applications, we consistently observe faster convergence, meaning our implementation improves throughput without significantly sacrificing algorithm performance in terms of reward and convergence rate.

\subsection{User Productivity}
\begin{table}[h]
\centering
\caption{User Productivity}
\label{tab:productivity}
\begin{threeparttable}
\begin{tabular}{ccc}
\hline
Algorithms       & DQN             & DDPG             \\ \hline
\multicolumn{1}{c|}{User code}                                                                                     & $\sim$75 lines  & $\sim$110 lines  \\
\multicolumn{1}{c|}{Development effort \tnote{$\triangleleft$}}    & $\sim$7 minutes & $\sim$10 minutes \\
\multicolumn{1}{c|}{Productivity across platforms ($CD$)} & $\sim$0.06      & $\sim$0.04       \\ \hline
\end{tabular}
\begin{tablenotes}\footnotesize
\item[$\triangleleft$] The compiling time for the FPGA image is excluded.
\end{tablenotes}
\end{threeparttable}
\end{table}
For a quick assessment of programmability, we enlisted a graduate student to implement two algorithms using PEARL. Table \ref{tab:productivity} quantifies the development effort involved. 
Note that we exclude FPGA image compilation time in Table \ref{tab:productivity} (as consistent with established practice \cite{chen2021thundergp}), since it is an integral part of the oneAPI workflow, and is not a step directly specified by PEARL users.
In addition to illustrating the effort required for developing a specific algorithm, we also present the Code Divergence ($CD$) to demonstrate productivity differences between development on two distinct platforms ($Server_{CGF}$ and $Server_{CG}$). $CD$ between two platforms, $i$ and $j$, is computed using the formula $CD=1-\frac{\left|c_i \cap c_j\right|}{\left|c_i \cup c_j\right|}$ \cite{pennycook2021navigating}, where $c$ represents the lines of user code. The $CD$ value falls within the range [0,1]: a value of 0 indicates that a ``single-source" code can be shared between both platforms, while a value of 1 implies that the user code is entirely different for the two platforms. In our case, $CD$ is close to 0, as the only required changes when porting to different devices involve modifying the paths to input files.

Overall, DRL application development through training in simulation is for tuning the best model and set of hyper-parameters before physical deployment. This requires repeated rounds of testing with different algorithms, hyper-parameters, and environmental scenarios to ensure the optimal reward and reliability of the AI agent. On state-of-the-art data-centers, it is unrealistic for application users to hand-tune each round of testing.
With PEARL, developers write only dozens of lines of code for generating the accelerated training-in-simulation implementation within minutes, significantly reducing the development effort. 
This acceleration of the development process enable DRL application users to efficiently explore a broader design space, leading to more robust AI agents with faster development cycles.
\vspace{-4pt}
\section{Conclusion  \& Future Work}
\vspace{-4pt}

We presented PEARL, a toolkit to facilitate productive development of high-performance DRL on heterogeneous platforms. 
Our experiments demonstrate performance gains across various algorithms, benchmark environments, and platforms. 
While PEARL unlocks the power of heterogeneous computing for a wide range of DRL algorithms, there are several research avenues for future exploration. First, as disaggregated data centers become popular, scaling and optimizing the distribution of each DRL primitive across multiple disaggregated heterogeneous nodes is a noteworthy research direction. Additionally, emerging DRL training functions can involve complex interactions among multiple DNN models; developing general-purpose tools based on intermediate task graph representations for mapping custom-defined training algorithms onto heterogeneous hardware holds significant promise. 


\newpage
\bibliographystyle{IEEEtran}
\bibliography{bib/ref, bib/chi_bib}

\begin{thebibliography}{10}
\providecommand{\url}[1]{#1}
\csname url@samestyle\endcsname
\providecommand{\newblock}{\relax}
\providecommand{\bibinfo}[2]{#2}
\providecommand{\BIBentrySTDinterwordspacing}{\spaceskip=0pt\relax}
\providecommand{\BIBentryALTinterwordstretchfactor}{4}
\providecommand{\BIBentryALTinterwordspacing}{\spaceskip=\fontdimen2\font plus
\BIBentryALTinterwordstretchfactor\fontdimen3\font minus \fontdimen4\font\relax}
\providecommand{\BIBforeignlanguage}[2]{{%
\expandafter\ifx\csname l@#1\endcsname\relax
\typeout{** WARNING: IEEEtran.bst: No hyphenation pattern has been}%
\typeout{** loaded for the language `#1'. Using the pattern for}%
\typeout{** the default language instead.}%
\else
\language=\csname l@#1\endcsname
\fi
#2}}
\providecommand{\BIBdecl}{\relax}
\BIBdecl

\bibitem{chatzilygeroudis2017black}
K.~Chatzilygeroudis, R.~Rama, R.~Kaushik, D.~Goepp, V.~Vassiliades, and J.-B. Mouret, ``Black-box data-efficient policy search for robotics,'' in \emph{2017 IEEE/RSJ International Conference on Intelligent Robots and Systems (IROS)}.\hskip 1em plus 0.5em minus 0.4em\relax IEEE, 2017, pp. 51--58.

\bibitem{vinyals2019alphastar}
O.~Vinyals, I.~Babuschkin, J.~Chung, M.~Mathieu, M.~Jaderberg, W.~M. Czarnecki, A.~Dudzik, A.~Huang, P.~Georgiev, R.~Powell \emph{et~al.}, ``Alphastar: Mastering the real-time strategy game starcraft ii,'' \emph{DeepMind blog}, vol.~2, 2019.

\bibitem{zhang2021parallel}
C.~Zhang, S.~R. Kuppannagari, and V.~K. Prasanna, ``Parallel actors and learners: A framework for generating scalable rl implementations,'' in \emph{2021 IEEE 28th International Conference on High Performance Computing, Data, and Analytics (HiPC)}.\hskip 1em plus 0.5em minus 0.4em\relax IEEE, 2021, pp. 1--10.

\bibitem{ray_rllib}
\BIBentryALTinterwordspacing
E.~Liang, R.~Liaw, R.~Nishihara, P.~Moritz, R.~Fox, J.~Gonzalez, K.~Goldberg, and I.~Stoica, ``Ray rllib: {A} composable and scalable reinforcement learning library,'' \emph{CoRR}, vol. abs/1712.09381, 2017. [Online]. Available: \url{http://arxiv.org/abs/1712.09381}
\BIBentrySTDinterwordspacing

\bibitem{cho2019fa3c}
H.~Cho, P.~Oh, J.~Park, W.~Jung, and J.~Lee, ``Fa3c: Fpga-accelerated deep reinforcement learning,'' in \emph{Proceedings of the Twenty-Fourth International Conference on Architectural Support for Programming Languages and Operating Systems}.\hskip 1em plus 0.5em minus 0.4em\relax ACM, 2019, pp. 499--513.

\bibitem{meng2020accelerating}
Y.~Meng, S.~Kuppannagari, and V.~Prasanna, ``Accelerating proximal policy optimization on cpu-fpga heterogeneous platforms,'' in \emph{2020 IEEE 28th Annual International Symposium on Field-Programmable Custom Computing Machines (FCCM)}.\hskip 1em plus 0.5em minus 0.4em\relax IEEE, 2020, pp. 19--27.

\bibitem{devcloud}
\BIBentryALTinterwordspacing
``Intel heterogeneous devcloud.'' [Online]. Available: \url{https://devcloud.intel.com/oneapi/}
\BIBentrySTDinterwordspacing

\bibitem{yasar2022pgabb}
A.~Yasar, S.~Rajamanickam, J.~W. Berry, and U.~V. Catalyurek, ``Pgabb: A block-based graph processing framework for heterogeneous platforms,'' \emph{arXiv preprint arXiv:2209.04541}, 2022.

\bibitem{van2012accelerating}
B.~Van~Essen, C.~Macaraeg, M.~Gokhale, and R.~Prenger, ``Accelerating a random forest classifier: Multi-core, gp-gpu, or fpga?'' in \emph{2012 IEEE 20th International Symposium on Field-Programmable Custom Computing Machines}.\hskip 1em plus 0.5em minus 0.4em\relax IEEE, 2012, pp. 232--239.

\bibitem{pennycook2021navigating}
S.~J. Pennycook, J.~D. Sewall, D.~W. Jacobsen, T.~Deakin, and S.~McIntosh-Smith, ``Navigating performance, portability, and productivity,'' \emph{Computing in Science \& Engineering}, vol.~23, no.~5, pp. 28--38, 2021.

\bibitem{oneapi}
\BIBentryALTinterwordspacing
Intel, ``Intel oneapi.'' [Online]. Available: \url{https://www.intel.com/content/www/us/en/developer/tools/oneapi/overview.html}
\BIBentrySTDinterwordspacing

\bibitem{dpcpp}
``Dpc++,'' https://www.intel.com/content/www/us/en/developer/tools/ oneapi/data-parallel-c-plus-plus.html.

\bibitem{sycl}
\BIBentryALTinterwordspacing
``Sycl.'' [Online]. Available: \url{https://registry.khronos.org/SYCL}
\BIBentrySTDinterwordspacing

\bibitem{dqn}
\BIBentryALTinterwordspacing
V.~Mnih, K.~Kavukcuoglu, D.~Silver, A.~Graves, I.~Antonoglou, D.~Wierstra, and M.~A. Riedmiller, ``Playing atari with deep reinforcement learning,'' \emph{CoRR}, vol. abs/1312.5602, 2013. [Online]. Available: \url{http://arxiv.org/abs/1312.5602}
\BIBentrySTDinterwordspacing

\bibitem{ddpg}
T.~P. Lillicrap, J.~J. Hunt, A.~Pritzel, N.~M.~O. Heess, T.~Erez, Y.~Tassa, D.~Silver, and D.~Wierstra, ``Continuous control with deep reinforcement learning,'' \emph{CoRR}, vol. abs/1509.02971, 2016.

\bibitem{meng2020efficiently}
Y.~Meng, Y.~Yang, S.~Kuppannagari, R.~Kannan, and V.~Prasanna, ``How to efficiently train your ai agent? characterizing and evaluating deep reinforcement learning on heterogeneous platforms,'' in \emph{2020 IEEE High Performance Extreme Computing Conference (HPEC)}.\hskip 1em plus 0.5em minus 0.4em\relax IEEE, 2020, pp. 1--7.

\bibitem{schaul2015prioritized}
T.~Schaul, J.~Quan, I.~Antonoglou, and D.~Silver, ``Prioritized experience replay,'' \emph{arXiv preprint arXiv:1511.05952}, 2015.

\bibitem{hessel2018rainbow}
M.~Hessel, J.~Modayil, H.~Van~Hasselt, T.~Schaul, G.~Ostrovski, W.~Dabney, D.~Horgan, B.~Piot, M.~Azar, and D.~Silver, ``Rainbow: Combining improvements in deep reinforcement learning,'' in \emph{Proceedings of the AAAI conference on artificial intelligence}, vol.~32, no.~1, 2018.

\bibitem{zhang2023framework}
C.~Zhang, Y.~Meng, and V.~Prasanna, ``A framework for mapping drl algorithms with prioritized replay buffer onto heterogeneous platforms,'' \emph{IEEE Transactions on Parallel and Distributed Systems}, 2023.

\bibitem{sgd}
H.~Robbins and S.~Monro, ``A stochastic approximation method,'' \emph{Annals of Mathematical Statistics}, vol.~22, pp. 400--407, 1951.

\bibitem{hacc}
\BIBentryALTinterwordspacing
``Amd heterogeneous accelerated compute clusters.'' [Online]. Available: \url{https://www.amd-haccs.io/}
\BIBentrySTDinterwordspacing

\bibitem{barba2021scientific}
L.~A. Barba, A.~Klockner, P.~Ramachandran, and R.~Thomas, ``Scientific computing with python on high-performance heterogeneous systems,'' \emph{Computing in Science \& Engineering}, vol.~23, no.~04, pp. 5--7, 2021.

\bibitem{apex}
\BIBentryALTinterwordspacing
D.~Horgan, J.~Quan, D.~Budden, G.~Barth{-}Maron, M.~Hessel, H.~van Hasselt, and D.~Silver, ``Distributed prioritized experience replay,'' \emph{CoRR}, vol. abs/1803.00933, 2018. [Online]. Available: \url{http://arxiv.org/abs/1803.00933}
\BIBentrySTDinterwordspacing

\bibitem{meng2022fpga}
Y.~Meng, C.~Zhang, and V.~Prasanna, ``Fpga acceleration of deep reinforcement learning using on-chip replay management,'' in \emph{Proceedings of the 19th ACM International Conference on Computing Frontiers}, 2022, pp. 40--48.

\bibitem{sac}
\BIBentryALTinterwordspacing
T.~Haarnoja, A.~Zhou, P.~Abbeel, and S.~Levine, ``Soft actor-critic: Off-policy maximum entropy deep reinforcement learning with a stochastic actor,'' \emph{CoRR}, vol. abs/1801.01290, 2018. [Online]. Available: \url{http://arxiv.org/abs/1801.01290}
\BIBentrySTDinterwordspacing

\bibitem{td3}
S.~Fujimoto, H.~V. Hoof, and D.~Meger, ``Addressing function approximation error in actor-critic methods,'' \emph{ArXiv}, vol. abs/1802.09477, 2018.

\bibitem{openai_gym}
G.~Brockman, V.~Cheung, L.~Pettersson, J.~Schneider, J.~Schulman, J.~Tang, and W.~Zaremba, ``Openai gym,'' 2016.

\bibitem{pytorch}
\BIBentryALTinterwordspacing
A.~Paszke, S.~Gross, F.~Massa, A.~Lerer, J.~Bradbury, G.~Chanan, T.~Killeen, Z.~Lin, N.~Gimelshein, L.~Antiga, A.~Desmaison, A.~Kopf, E.~Yang, Z.~DeVito, M.~Raison, A.~Tejani, S.~Chilamkurthy, B.~Steiner, L.~Fang, J.~Bai, and S.~Chintala, ``Pytorch: An imperative style, high-performance deep learning library,'' in \emph{Advances in Neural Information Processing Systems 32}, H.~Wallach, H.~Larochelle, A.~Beygelzimer, F.~d\textquotesingle Alch\'{e}-Buc, E.~Fox, and R.~Garnett, Eds.\hskip 1em plus 0.5em minus 0.4em\relax Curran Associates, Inc., 2019, pp. 8024--8035. [Online]. Available: \url{http://papers.neurips.cc/paper/9015-pytorch-an-imperative-style-high-performance-deep-learning-library.pdf}
\BIBentrySTDinterwordspacing

\bibitem{pybind11}
W.~Jakob, J.~Rhinelander, and D.~Moldovan, ``pybind11 — seamless operability between c++11 and python,'' 2016, https://github.com/pybind/pybind11.

\bibitem{autorun}
\BIBentryALTinterwordspacing
Intel, ``Sycl c++ autorun feature for fpga.'' [Online]. Available: \url{https://github.com/oneapi-src/oneAPI-samples/tree/master/DirectProgramming/C%2B%2BSYCL_FPGA/Tutorials/DesignPatterns/autorun}
\BIBentrySTDinterwordspacing

\bibitem{torchintel}
\BIBentryALTinterwordspacing
``Intel extension for pytorch.'' [Online]. Available: \url{https://github.com/intel/intel-extension-for-pytorch}
\BIBentrySTDinterwordspacing

\bibitem{stable-baselines3}
\BIBentryALTinterwordspacing
A.~Raffin, A.~Hill, A.~Gleave, A.~Kanervisto, M.~Ernestus, and N.~Dormann, ``Stable-baselines3: Reliable reinforcement learning implementations,'' \emph{Journal of Machine Learning Research}, vol.~22, no. 268, pp. 1--8, 2021. [Online]. Available: \url{http://jmlr.org/papers/v22/20-1364.html}
\BIBentrySTDinterwordspacing

\bibitem{chen2021thundergp}
X.~Chen, H.~Tan, Y.~Chen, B.~He, W.-F. Wong, and D.~Chen, ``Thundergp: Hls-based graph processing framework on fpgas,'' in \emph{The 2021 ACM/SIGDA International Symposium on Field-Programmable Gate Arrays}, 2021, pp. 69--80.

\end{thebibliography}

\end{document}